\pdfoutput=1
\documentclass[12pt,a4paper]{article}
\usepackage{ifthen} % for conditional statements
\usepackage{wrapfig}
\usepackage{float}
\newfloat{lstfloat}{htbp}{lop}
\floatname{lstfloat}{Listing}

\newboolean{pdflatex}
\setboolean{pdflatex}{true} % False for eps figures 

\newboolean{articletitles}
\setboolean{articletitles}{true} % False removes titles in references

\newboolean{uprightparticles}
\setboolean{uprightparticles}{false} %True for upright particle symbols

\def\paperauthors{
  Abhijit Mathad$^{1,2}$$^\dagger$,
  Martina Ferrillo$^1$,
  Sacha Barr\'e$^{2,3}$,
  Patrick Koppenburg$^4$,
  Patrick Owen$^1$,
  Gerhard Raven$^{4,5}$,
  Eduardo Rodrigues$^6$,
  Nicola Serra$^1$
  } % Leave as is for PAPER, CONF and FIGURE
\def\paperasciititle{FunTuple: A new N-tuple component for offline data processing at the LHCb experiment} % Latex formatted title
\def\papertitle{FunTuple: A new N-tuple component for offline data processing at the LHCb experiment} % Latex formatted title
\def\paperkeywords{{High Energy Physics}, {LHCb}, {nTuples}, {Offline data processing}} % Comma separated list
\def\papercopyright{\the\year\ CERN for the benefit of the LHCb collaboration} % new since 9/Apr/2018
\def\paperlicence{CC BY 4.0 licence}
\def\paperlicenceurl{https://creativecommons.org/licenses/by/4.0/}

\usepackage[top=1in, bottom=1.25in, left=1in, right=1in]{geometry}

\usepackage{microtype}
\usepackage{lineno}  % for line numbering during review
\usepackage{xspace} % To avoid problems with missing or double spaces after
\usepackage{caption} %these three command get the figure and table captions automatically small

\usepackage{graphicx}  % to include figures (can also use other packages)
\usepackage{color}
\usepackage{colortbl}
\usepackage{amsmath} % Adds a large collection of math symbols
\usepackage{amssymb}
\usepackage{amsfonts}
\usepackage{upgreek} % Adds in support for greek letters in roman typeset

\newcommand*\patchAmsMathEnvironmentForLineno[1]{%
\expandafter\let\csname old#1\expandafter\endcsname\csname #1\endcsname
\expandafter\let\csname oldend#1\expandafter\endcsname\csname
end#1\endcsname
 \renewenvironment{#1}%
   {\linenomath\csname old#1\endcsname}%
   {\csname oldend#1\endcsname\endlinenomath}%
}
\newcommand*\patchBothAmsMathEnvironmentsForLineno[1]{%
  \patchAmsMathEnvironmentForLineno{#1}%
  \patchAmsMathEnvironmentForLineno{#1*}%
}
\AtBeginDocument{%
\patchBothAmsMathEnvironmentsForLineno{equation}%
\patchBothAmsMathEnvironmentsForLineno{align}%
\patchBothAmsMathEnvironmentsForLineno{flalign}%
\patchBothAmsMathEnvironmentsForLineno{alignat}%
\patchBothAmsMathEnvironmentsForLineno{gather}%
\patchBothAmsMathEnvironmentsForLineno{multline}%
\patchBothAmsMathEnvironmentsForLineno{eqnarray}%
}

\usepackage{hyperxmp}

\usepackage[pdftex,
            pdfauthor={\paperauthors},
            pdftitle={\paperasciititle},
            pdfkeywords={\paperkeywords},
            pdfcopyright={Copyright (C) \papercopyright},
            pdflicenseurl={\paperlicenceurl}]{hyperref}
\usepackage[colorinlistoftodos,textsize=scriptsize]{todonotes}

\usepackage[bottom,flushmargin,hang,multiple]{footmisc}

\usepackage[all]{hypcap} % Internal hyperlinks to floats.

\usepackage{xspace} 
\usepackage{upgreek}

\def\lhcb   {\mbox{LHCb}\xspace}

\def\MagUp {\mbox{\em Mag\kern -0.05em Up}\xspace}

\ifthenelse{\boolean{uprightparticles}}%
{

 \def\Pmu         {\ensuremath{\upmu}\xspace}

 \def\Ppi         {\ensuremath{\uppi}\xspace}

 \def\Ppsi        {\ensuremath{\uppsi}\xspace}

 \def\PDelta      {\ensuremath{\Delta}\xspace}                 
 \def\PXi         {\ensuremath{\Xi}\xspace}                 
 \def\PLambda     {\ensuremath{\Lambda}\xspace}                 
 \def\PSigma      {\ensuremath{\Sigma}\xspace}                 
 \def\POmega      {\ensuremath{\Omega}\xspace}                 
 \def\PUpsilon    {\ensuremath{\Upsilon}\xspace}
 \let\oldPi\Pi
 \def\PPi         {\ensuremath{\oldPi}\xspace}

 \def\PB      {\ensuremath{\mathrm{B}}\xspace}                 
                  
 \def\PD      {\ensuremath{\mathrm{D}}\xspace}

 \def\PJ      {\ensuremath{\mathrm{J}}\xspace}                 
 \def\PK      {\ensuremath{\mathrm{K}}\xspace}

 \def\Pi      {\ensuremath{\mathrm{i}}\xspace}

 \def\Ps      {\ensuremath{\mathrm{s}}\xspace}

 \def\thebaroffset{0.0em}
}
{

 \def\Pmu         {\ensuremath{\mu}\xspace}

 \def\Ppi         {\ensuremath{\pi}\xspace}

 \def\Ppsi        {\ensuremath{\psi}\xspace}                 
                  
 \mathchardef\PDelta="7101
 \mathchardef\PXi="7104
 \mathchardef\PLambda="7103
 \mathchardef\PSigma="7106
 \mathchardef\POmega="710A
 \mathchardef\PUpsilon="7107
 \mathchardef\PPi="7105
                  
 \def\PB      {\ensuremath{B}\xspace}                 
                  
 \def\PD      {\ensuremath{D}\xspace}

 \def\PJ      {\ensuremath{J}\xspace}                 
 \def\PK      {\ensuremath{K}\xspace}

 \def\Pi      {\ensuremath{i}\xspace}

 \def\Ps      {\ensuremath{s}\xspace}

 \def\thebaroffset{0.18em}
}
\newcommand{\offsetoverline}[2][\thebaroffset]{\kern #1\overline{\kern -#1 #2}}%

\makeatletter
\ifcase \@ptsize \relax% 10pt
  \newcommand{\miniscule}{\@setfontsize\miniscule{4}{5}}% \tiny: 5/6
\or% 11pt
  \newcommand{\miniscule}{\@setfontsize\miniscule{5}{6}}% \tiny: 6/7
\or% 12pt
  \newcommand{\miniscule}{\@setfontsize\miniscule{5}{6}}% \tiny: 6/7
\fi
\makeatother

\DeclareRobustCommand{\optbar}[1]{\shortstack{{\miniscule (\rule[.5ex]{1.25em}{.18mm})}
  \\ [-.7ex] $#1$}}

\def\mup        {{\ensuremath{\Pmu^+}}\xspace}
\def\mun        {{\ensuremath{\Pmu^-}}\xspace} % muon negative (\mum is taken)

\def\squark    {{\ensuremath{\Ps}}\xspace}

\def\pion   {{\ensuremath{\Ppi}}\xspace}

\def\pip    {{\ensuremath{\pion^+}}\xspace}

\def\kaon    {{\ensuremath{\PK}}\xspace}
\def\KorKbar {\kern \thebaroffset\optbar{\kern -\thebaroffset \PK}{}\xspace}

\def\Kp      {{\ensuremath{\kaon^+}}\xspace}
\def\Km      {{\ensuremath{\kaon^-}}\xspace}

\def\D       {{\ensuremath{\PD}}\xspace}

\def\DorDbar {\kern \thebaroffset\optbar{\kern -\thebaroffset \PD}\xspace}

\def\Dp      {{\ensuremath{\D^+}}\xspace}
\def\Dm      {{\ensuremath{\D^-}}\xspace}

\def\DpDm    {\ensuremath{\Dp {\kern -0.16em \Dm}}\xspace}

\def\B       {{\ensuremath{\PB}}\xspace}

\def\BorBbar {\kern \thebaroffset\optbar{\kern -\thebaroffset \PB}\xspace}

\def\Bd      {{\ensuremath{\B^0}}\xspace}

\def\BdorBdbar {\kern \thebaroffset\optbar{\kern -\thebaroffset \Bd}\xspace}
\def\Bu      {{\ensuremath{\B^+}}\xspace}
\def\Bub     {{\ensuremath{\B^-}}\xspace}
\def\Bp      {{\ensuremath{\Bu}}\xspace}
\def\Bm      {{\ensuremath{\Bub}}\xspace}

\def\Bs      {{\ensuremath{\B^0_\squark}}\xspace}

\def\BsorBsbar {\kern \thebaroffset\optbar{\kern -\thebaroffset \Bs}\xspace}

\def\jpsi     {{\ensuremath{{\PJ\mskip -3mu/\mskip -2mu\Ppsi{(1S)}}}}\xspace}

\def\Y#1S{\ensuremath{\PUpsilon{(#1S)}}\xspace}

\def\LorLbar     {\kern \thebaroffset\optbar{\kern -\thebaroffset \PLambda}\xspace}

\newcommand{\decay}[2]{\ensuremath{#1\!\to #2}\xspace} 

\def\to                 {\ensuremath{\rightarrow}\xspace}

\def\AT#1     {\ensuremath{A_{\mathrm{T}}^{#1}}\xspace}           % 2

\def\C#1      {\ensuremath{\mathcal{C}_{#1}}\xspace}                       % 9
\def\Cp#1     {\ensuremath{\mathcal{C}_{#1}^{'}}\xspace}                    % 7
\def\Ceff#1   {\ensuremath{\mathcal{C}_{#1}^{\mathrm{(eff)}}}\xspace}        % 9  
\def\Cpeff#1  {\ensuremath{\mathcal{C}_{#1}^{'\mathrm{(eff)}}}\xspace}       % 7
\def\Ope#1    {\ensuremath{\mathcal{O}_{#1}}\xspace}                       % 2
\def\Opep#1   {\ensuremath{\mathcal{O}_{#1}^{'}}\xspace}                    % 7

\newcommand{\aunit}[1]{\ensuremath{\text{\,#1}}}       
\newcommand{\tev}{\aunit{Te\kern -0.1em V}\xspace}
\newcommand{\gev}{\aunit{Ge\kern -0.1em V}\xspace}
\newcommand{\mev}{\aunit{Me\kern -0.1em V}\xspace}
\newcommand{\kev}{\aunit{ke\kern -0.1em V}\xspace}
\newcommand{\ev}{\aunit{e\kern -0.1em V}\xspace}
 
\newcommand{\mevc}{\ensuremath{\aunit{Me\kern -0.1em V\!/}c}\xspace}
\newcommand{\gevc}{\ensuremath{\aunit{Ge\kern -0.1em V\!/}c}\xspace}
\newcommand{\mevcc}{\ensuremath{\aunit{Me\kern -0.1em V\!/}c^2}\xspace}
\newcommand{\gevcc}{\ensuremath{\aunit{Ge\kern -0.1em V\!/}c^2}\xspace}
\def\fb   {\ensuremath{\aunit{fb}}\xspace}
\def\invfb   {\ensuremath{\fb^{-1}}\xspace}

\def\gsim{{~\raise.15em\hbox{$>$}\kern-.85em
          \lower.35em\hbox{$\sim$}~}\xspace}
\def\lsim{{~\raise.15em\hbox{$<$}\kern-.85em
          \lower.35em\hbox{$\sim$}~}\xspace}

\def\cpp        {\mbox{\textsc{C\raisebox{0.1em}{{\footnotesize{++}}}}}\xspace}
\def\tell1  {TELL1\xspace}
\def\ukl1   {UKL1\xspace}

\def\CC{{C\nolinebreak[4]\hspace{-.05em}\raisebox{.4ex}{\tiny\bf ++}}} % Add in the predefined LHCb symbols

\usepackage{cite} % Allows for ranges in citations
\usepackage{mciteplus}
\usepackage{longtable} % only for template; not usually to be used in PAPERs
\usepackage{listings}
\usepackage{graphicx}
\usepackage{float}
\usepackage{url}
\usepackage{tikz}
\usepackage{subcaption}

\definecolor{dkgreen}{rgb}{0,0.6,0}
\definecolor{gray}{rgb}{0.5,0.5,0.5}
\definecolor{mauve}{rgb}{0.58,0,0.82}
\definecolor{backcolour}{rgb}{0.95,0.95,0.92}
\definecolor{cobalt}{rgb}{0.0, 0.28, 0.67}
\definecolor{darkpowderblue}{rgb}{0.0, 0.2, 0.6}
\definecolor{darkmidnightblue}{rgb}{0.0, 0.2, 0.4}
\lstdefinestyle{mystyle}{frame=tb,
    language=C++,
    aboveskip=3mm,
    belowskip=3mm,
    showstringspaces=false,,
    columns=flexible,
    basicstyle={\small\ttfamily},
    numbers=left,
    numbersep=5pt,
    backgroundcolor=\color{backcolour},
    numberstyle=\scriptsize\color{gray},
    keywordstyle=\color{blue},
    morekeywords={Node, Tree, Arrow},
    keywordstyle=[2]\color{dkgreen},
    morekeywords=[2]{PASSED},
    commentstyle=\color{dkgreen},
    stringstyle=\color{mauve},
    breaklines=true,
    breakatwhitespace=true,
    tabsize=2
}

\begin{document}

%%%%%%%%%%%%%%%%%%%%%%%%%
%%%%% Title     %%%%%%%%%
%%%%%%%%%%%%%%%%%%%%%%%%%
\renewcommand{\thefootnote}{\fnsymbol{footnote}}
\setcounter{footnote}{1}

% %%%%%%% CHOOSE TITLE PAGE--------
%\onecolumn
% ===============================================================================
% Purpose: LHCb-INT Note title page template
% Author: P. Koppenburg
% Created on: 2015-05-18
% ===============================================================================

%%%%%%%%%%%%%%%%%%%%%%%%%
%%%%%  TITLE PAGE  %%%%%%
%%%%%%%%%%%%%%%%%%%%%%%%%
\begin{titlepage}

% Header ---------------------------------------------------
\vspace*{-1.5cm}

\noindent
\begin{tabular*}{\linewidth}{lc@{\extracolsep{\fill}}r@{\extracolsep{0pt}}}
\ifthenelse{\boolean{pdflatex}}% Logo format choice
%{\vspace*{-1.2cm}\mbox{\!\!\!\includegraphics[width=.14\textwidth]{figs/lhcb-logo.pdf}} & &}%
%{\vspace*{-1.2cm}\mbox{\!\!\!\includegraphics[width=.12\textwidth]{figs/lhcb-logo.eps}} & &}
 \\
 %& & %LHCb-INT-2022-005 
 \\  % ID 
 & & \today \\ % Date - Can also hardwire e.g.: 23 March 2010
 & & \\
\hline
\end{tabular*}

\vspace*{4.0cm}

% Title --------------------------------------------------
{\normalfont\bfseries\boldmath\huge
\begin{center}
% DO NOT EDIT HERE. Instead edit macro in main.tex to keep metadata correct
  \papertitle
\end{center}
}

\vspace*{2.0cm}

% Authors -------------------------------------------------
\begin{center}
% If changing to list here, make pdfauthors in main.tex a comma
% separated list with the same names. Otherwise metadata in file will be wrong.
\paperauthors
\bigskip\\
{\normalfont\itshape\footnotesize
$^1$ University of Z\"urich, Z\"urich, Switzerland\\
$^2$ European Organization for Nuclear Research (CERN), Geneva, Switzerland\\
$^3$ The University of Manchester, Manchester, United Kingdom\\
$^4$ Nikhef National Institute for Subatomic Physics, Amsterdam, Netherlands\\
$^5$ VU University Amsterdam, Amsterdam, Netherlands\\
$^6$ Oliver Lodge Laboratory, University of Liverpool, Liverpool, United Kingdom\\
\vspace{0.5cm}
$^\dagger$ Contact author: \href{mailto:amathad@cern.ch}{amathad@cern.ch}\\
\centerline{Keywords: High-energy-physics, LHCb experiment, Data Processing and Offline Analysis}
}

\end{center}

\vspace{\fill}

% Abstract -----------------------------------------------
\begin{abstract}
The offline software framework of the \lhcb experiment has undergone a significant overhaul to tackle the data processing challenges that will arise in the upcoming Run 3 and Run 4 of the Large Hadron Collider. 
This paper introduces \texttt{FunTuple}, a novel component developed for offline data processing within the \lhcb experiment.
This component enables the computation and storage of 
a diverse range of observables for both reconstructed and simulated events by 
leveraging on the tools initially developed for the trigger system. 
This feature is crucial for ensuring
consistency between trigger-computed and offline-analysed observables.
The component and its tool suite offer users flexibility to customise stored observables, and its reliability is 
validated through a full-coverage set of rigorous unit tests.
This paper comprehensively explores \texttt{FunTuple}'s design, interface, interaction with other algorithms, and 
its role in facilitating offline data processing for the \lhcb experiment for the next decade and beyond.
\end{abstract}

\vspace*{1.0cm}
\vspace{\fill}
{\footnotesize 
% Edit macro in main.tex to keep metadata correct
\centerline{\copyright~\papercopyright. \href{\paperlicenceurl}{\paperlicence}.}}
%\vspace{0.5cm}
\vspace*{2mm}
\end{titlepage}

\pagestyle{empty}  % no page number for the title 

%%%%%%%%%%%%%%%%%%%%%%%%%%%%%%%%
%%%%%  EOD OF TITLE PAGE  %%%%%%
%%%%%%%%%%%%%%%%%%%%%%%%%%%%%%%%

%%  empty page follows the title page ----
%\newpage
%\setcounter{page}{2}
%\mbox{~}

%\input{title-LHCb-ANA}
%\input{title-LHCb-CONF}
%\input{title-LHCb-FIGURE}
%\input{title-LHCb-PAPER}
%\twocolumn
% %%%%%%%%%%%%% ---------

\renewcommand{\thefootnote}{\arabic{footnote}}
\setcounter{footnote}{0}

%%%%%%%%%%%%%%%%%%%%%%%%%%%%%%%%
%%%%%  Table of Content   %%%%%%
%%%%%%%%%%%%%%%%%%%%%%%%%%%%%%%%
%%%% Uncomment if desired
%\tableofcontents
\cleardoublepage

%%%%%%%%%%%%%%%%%%%%%%%%%
%%%%% Main text %%%%%%%%%
%%%%%%%%%%%%%%%%%%%%%%%%%

\pagestyle{plain} % restore page numbers for the main text
\setcounter{page}{1}
\pagenumbering{arabic}

%% Uncomment during review phase. 
%% Comment before a final submission.
%\linenumbers

%% This is the main body
%% It is useful to have a single file so comments are not missed in overleaf.
\section{Introduction}
\label{sec:Introduction}

The \lhcb experiment,
located at Point 8 of the Large Hadron Collider (LHC)~\cite{Evans:2008zzb} at CERN,
is a forward-arm spectrometer designed to study the decays of beauty and charm hadrons~\cite{LHCb-DP-2008-001,LHCb-DP-2014-002}.
In the initial two runs of the LHC, during 2010--2018, the experiment (mainly) collected proton-proton collision data corresponding to a total integrated luminosity of 9\invfb.
As preparations intensify for Run 3~\footnote{The data collection from Run 3 of LHC is currently ongoing; however, the core developments emphasised in this paper transpired prior to its commencement.}, 
where the LHC's instantaneous luminosity is anticipated to surge by a factor of 5 compared to the preceding runs, 
the \lhcb experiment is poised to enhance its capabilities even further.
The upgraded detector~\cite{LHCb-DP-2022-002} and data acquisition system will allow for improved vertexing and trigger efficiency~\cite{LHCb-DP-2019-002}.
This enhancement facilitates the exploration of exceedingly rare decays~\cite{LHCb-PAPER-2021-008} while also facilitating the probing of deviations from Standard Model predictions with unparalleled precision~\cite{LHCb-PAPER-2020-002,LHCb-PAPER-2022-052,LHCb-PAPER-2022-039}.

The advent of Run 3 data acquisition presents significant hurdles for the \lhcb data processing framework. 
Notably, the data volume from \lhcb's Run 3 is projected to surge by over 15 times compared to prior runs~\cite{Skidmore:2022rza}. 
Consequently, management of petabytes of processed data and effectively incorporating distributed computing resources present significant challenges~\cite{LHCb-TDR-018,Tsaregorodtsev:2010zz}.
In light of these challenges, a comprehensive redesign of both the trigger and offline data processing pipelines is imperative~\cite{LHCb-TDR-018,Skidmore:2022rza}. 
This paper concentrates on the offline data processing pipeline, specifically highlighting the development of a new component called \texttt{FunTuple}~\cite{FunTupleRepo} facilitating analysis of Run 3 data and beyond.

In the initial LHC runs, \lhcb's trigger 
and offline reconstruction applications, \texttt{Moore}~\cite{Moore} 
and \texttt{Brunel}~\cite{BrunelPackage}, 
operated independently 
from the \texttt{DaVinci} application~\cite{DaVinci} employed for offline data processing. 
Besides executing offline event selection, the \texttt{DaVinci} application was used 
to process and store data for subsequent analysis.
This task was accomplished via the \texttt{DecayTreeTuple} algorithm~\cite{Analysis}, \footnote{There were also alternative Python based algorithms 
like \texttt{Bender}~\cite{Belyaev:2004eda,Bender} for Run 1/2 data processing.} which recorded a specific set of observables into output files.
Firstly, due to the segregation of trigger and offline frameworks, the equivalence between trigger-computed observables and those analysed offline was not guaranteed. 
Secondly, users lacked the flexibility to customise the set of observables recorded, which is essential in light of the anticipated data volume surge for Run 3 and Run 4. 
Furthermore, as part of its strategy to tackle the forthcoming data processing challenges in Run 4 and beyond, the LHCb experiment plans to implement a new event model based on \textit{Structure of Arrays} (SoA), which will facilitate vectorised processing of data~\cite{Hennequin:2023tjg}.
Substantial enhancements were also made to
the trigger reconstruction algorithms that facilitated retirement of the \texttt{Brunel} package, which was responsible for offline reconstruction~\cite{Aaij:2019zbu,LHCb:2023hlw, Fitzpatrick:1670985,Reiss:2846414}. 
Consequently, the development of
new offline algorithms becomes imperative to accommodate these changes.

\begin{figure}[!hpt]
\centering
\includegraphics[width=\textwidth]{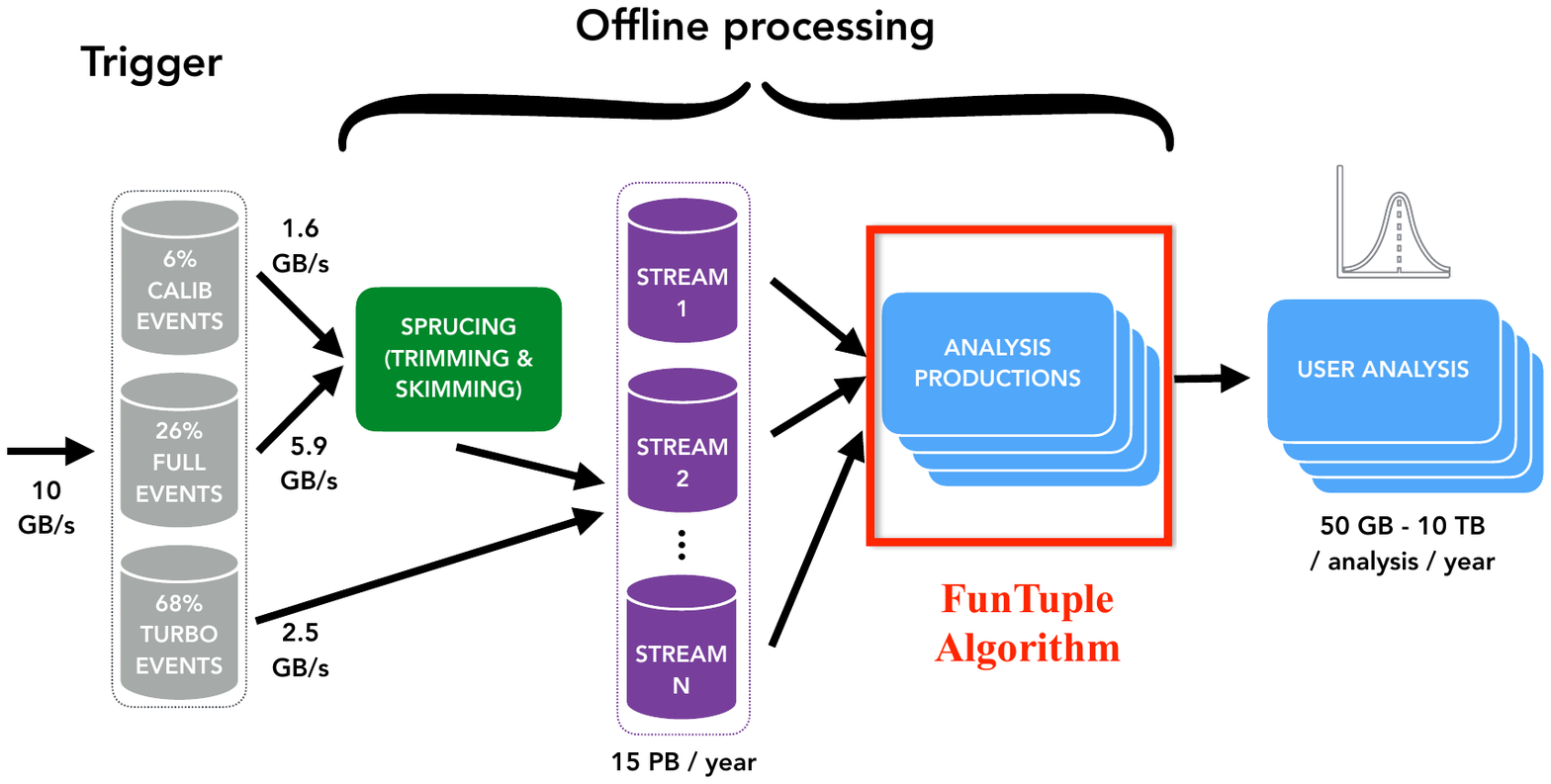}
\caption{Data flow diagram for Run 3 data processing showing the placement of the \texttt{FunTuple} component. Figure adapted from Ref.~\cite{LHCB-FIGURE-2020-016}.}
\label{fig:flow}
\end{figure}

To overcome these hurdles, a strategic choice was made to leverage 
tools developed for the trigger system within the offline software framework.
This led to the development of a new component, \texttt{FunTuple}~\cite{FunTupleRepo}, short for \textbf{Fun}ctional n\textbf{Tuple}, which is tailored for processing Run 3 and Run 4 data. 
The \texttt{FunTuple} component introduces enhancements to the previous workflow. Firstly, it guarantees the consistency between trigger-computed observables and those subjected to offline analysis. 
Secondly, FunTuple, along with all its dependencies, is entirely templated in C++, allowing it to support both legacy and upcoming event models planned for future LHC runs.
The templated design along with the SoA event model enables the component to leverage \textit{Single Instruction Multiple Data} (SIMD) vectorisation.
Lastly, it offers users the flexibility to efficiently tailor the list of recorded observables,
an important feature given the expected surge in data volume for Run 3 and Run 4.
This component is configured with a robust suite of tools designed for the second stage of the \lhcb trigger system, known as Throughput Oriented (\texttt{ThOr}) functors~\cite{ThOr-Functors,Nolte:2765896,Rec}. 
These functors are designed to deliver high-speed in the trigger's demanding throughput environment and are adept at computing topological and kinematic observables.
\texttt{FunTuple} utilises these functors to compute a diverse range of observables and writes 
a \texttt{TTree} in the \texttt{ROOT} N-tuple format.
\footnote{There are plans in the future to write \texttt{ROOT} \texttt{RNTuple}, which has been designed to address performance bottlenecks and shortcomings of \texttt{ROOT} current state of the art \texttt{TTree}~\cite{Lopez-Gomez:2022umr}.}
The {N-tuple} format is widely used in the High Energy Physics community to store flattened data in a tabular format~\cite{Brun:1997pa}.
Furthermore, the component's lightweight design ensures simplified maintenance and seamless knowledge transfer. 
As depicted in Fig.\ref{fig:flow}, the \texttt{FunTuple} component plays a central role, bridging the gap between the offline data processing stage (\texttt{Sprucing}) and the subsequent user analysis stages\cite{LHCB-FIGURE-2020-016}. In the \texttt{Sprucing} stage, the data is slimmed and skimmed before being saved to disk as part of the offline data processing workflow.
The placement of \texttt{FunTuple} underscores its critical role in \lhcb's analysis productions~\cite{AnalysisProd}, facilitating the storage of experiment-acquired data in a format suitable for subsequent offline analysis.

\section{Design and interface}
\label{sec:design}

\texttt{FunTuple} is a novel component integral to the \lhcb experiment's data processing infrastructure. 
It is a \cpp~\cite{cpp} class built upon the \texttt{Gaudi} functional framework~\cite{Barrand:2001ny}, and it offers a user-friendly \texttt{Python}~\cite{python} interface.
The flexibility of the \texttt{FunTuple} component lies in its templated design, allowing it to accommodate various types of input data. 
As a result, for Run 3, it is available in the three distinct flavours \texttt{FunTuple\_Particles}, \texttt{FunTuple\_MCParticles} and \texttt{FunTuple\_Event} hereafter described.

\begin{figure}[!hpt]
\centering
\includegraphics[width=\textwidth]{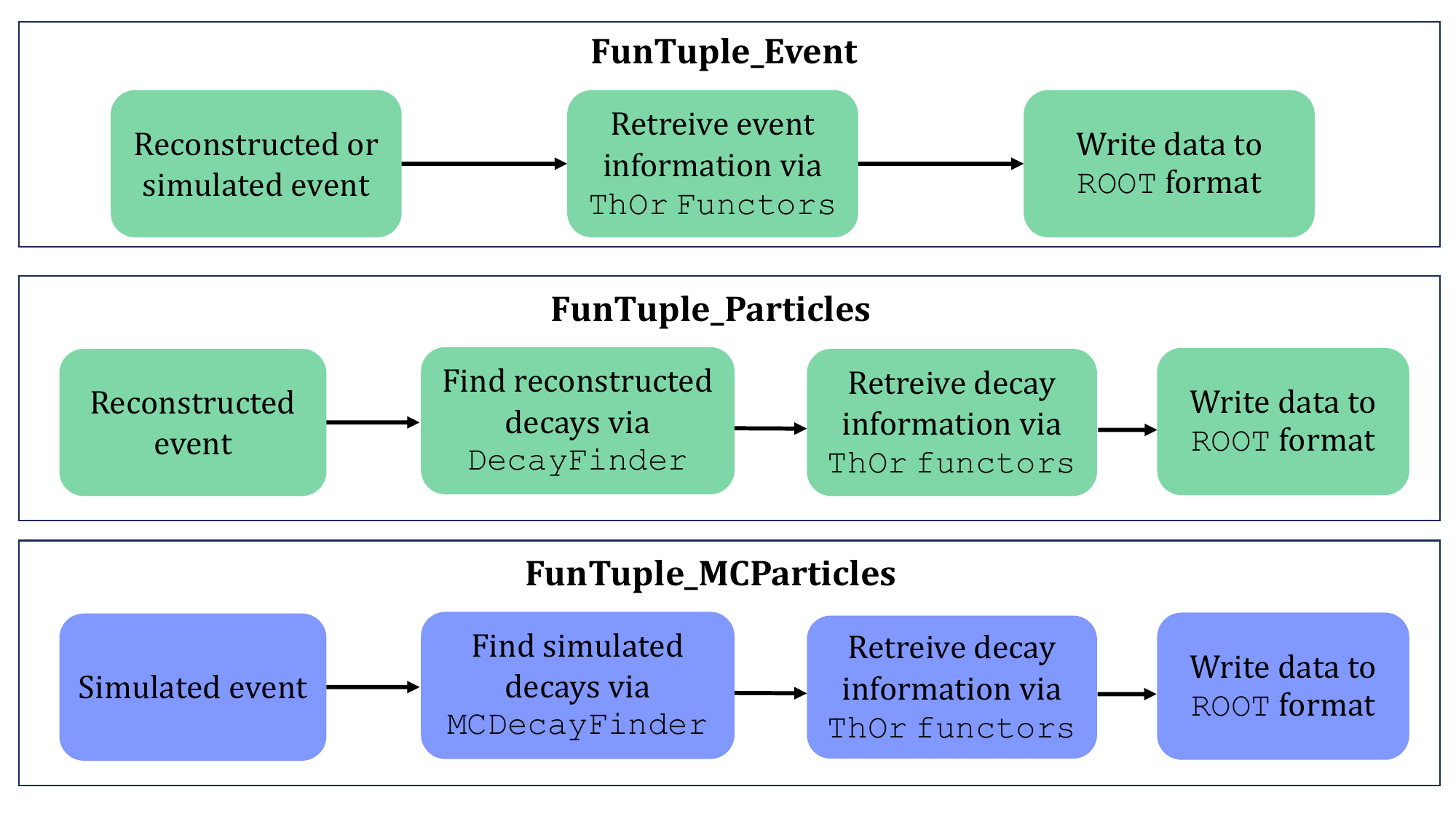}
\caption{Data flow diagram of the three flavours of \texttt{FunTuple} component.}
\label{fig:flow_tupling}
\end{figure}

The \texttt{FunTuple\_Event} component processes input data comprising of reconstructed or simulated events, where each event represents a single LHC bunch crossing. It acquires event-level information (for example the number of charged particles in the event),
using thread-safe \texttt{ThOr} functors that are specialised \cpp classes developed for utilisation in the second stage of the \lhcb trigger system~\cite{Li:2022tlq,Nolte:2765896,ThOr-Functors}. 
The component then stores this extracted information from \texttt{ThOr} functors in a \texttt{ROOT} {N-tuple}  file.
The \texttt{FunTuple\_Particles} component functions on reconstructed events and identifies specific reconstructed decays by utilising the decay-finding tool \texttt{DecayFinder}~\cite{Rec} explained in Section~\ref{sec:fields}. It further retrieves essential details regarding parent and children particles (for example magnitude of the transverse moment) through \texttt{ThOr} functors and records this information in a ROOT file.
Similarly, the \texttt{FunTuple\_MCParticles} component shares similarities with \texttt{FunTuple\_Particles}, but it processes simulated events instead, and captures information about simulated decays.
For an illustrative representation of the data flow encompassing these three approaches, refer to Fig.~\ref{fig:flow_tupling}. 
Each aspect of the data-flow diagram is further elaborated in the following sections.

The instantiation of the three flavours of the \texttt{FunTuple} component in \texttt{Python} is exemplified in Listings~\ref{lst:ftreco}--~\ref{lst:ftevt}.
As depicted, the user is required to provide the \texttt{name} and \texttt{tuple\_name}  attributes for all three flavours. 
The \texttt{name} attribute defines the component's name and the name of the corresponding \texttt{TDirectory} in the output ROOT file. 
On the other hand, the \texttt{tuple\_name} attribute defines the name of the \texttt{TTree} in the ROOT file.
The \texttt{fields} attribute can only be defined for \texttt{FunTuple\_Particles} and \texttt{FunTuple\_MCParticles} and is used to select specific decays within an event and define the corresponding \texttt{TBranches} in the output file.
For a detailed exploration of this attribute, see Section~\ref{sec:fields}.
The \texttt{variables} attribute is used to specify the observables to be computed for each event or decay.
In the case of \texttt{FunTuple\_Event}, only event-level observables can be defined. Conversely, for \texttt{FunTuple\_Particles} and \texttt{FunTuple\_MCParticles}, both 
decay-level and event-level observables can be specified. 
The latter is achieved by defining an optional \texttt{event\_variables} attribute.
It is worth noting that the \texttt{FunTuple} component automatically writes certain event information, such as the run and event numbers,\footnote{Both run and event numbers are used to uniquely identify an event in the LHC experiments.} to the output file by default. 
For a more comprehensive discussion on the \texttt{variables} attribute, refer to Section~\ref{sec:variables}. Finally, the \texttt{inputs} attribute refers to the Transient Event Store (TES) location, indicating the data pertaining to a given event cycle that will be processed.
Subsequently, the processed information is stored in the output ROOT file, which is further elaborated on in Section~\ref{sec:output}.

The \texttt{FunTuple} component also incorporates several essential counters to monitor the data processing. These counters include tracking the number of processed events, the count of non-empty events for each selected particle, and the tally of events with multiple candidates for each chosen particle. Upon completing the data processing, the results of these counters are displayed to the users.
To ensure effective error handling, the component employs a custom error handling class that inherits from the \texttt{StatusCode} class implemented in \texttt{Gaudi}. This custom implementation enables the component to raise specific exceptions in targeted scenarios. For example, if a particular \texttt{ThOr} functor encounters difficulties and cannot compute an observable for a given event, the component raises an exception to promptly notify the user of the issue.
Additionally, the \texttt{FunTuple} component takes measures to validate the input attributes both on the Python and \cpp sides, ensuring the correctness of the provided data.
Moreover, the development process includes the creation of several tests and examples, see Section~\ref{sec:tests_examples}.

\begin{lstfloat}
\begin{minipage}{.45\textwidth}
  \noindent
  \begin{lstlisting}[
  firstnumber=1,
  frame=tlrb, language=Python, caption={\texttt{FunTuple\_Particles} instance\vspace{0.05cm}}, label={lst:ftreco}]{Name}
# import FunTuple to run over reconstructed particles
from FunTuple import FunTuple_Particles

# define instance of FunTuple
data_tuple = FunTuple_Particles(
    name="TDirectoryName",
    tuple_name="TTreeName",
    fields=fields,
    variables=variables,
    event_variables=event_variables,
    inputs=reco_data_TES_location)
  \end{lstlisting}
  \end{minipage}
  \hfill
  \begin{minipage}{.45\textwidth}
  \begin{lstlisting}[
  firstnumber=1,
  frame=tlrb, language=Python, caption={\texttt{FunTuple\_MCParticles} instance\vspace{0.05cm}}, label={lst:ftmc}]{Name}
# import FunTuple to run over simulated particles
from FunTuple import FunTuple_MCParticles

# define instance of FunTuple
data_tuple = FunTuple_MCParticles(
    name="TDirectoryName",
    tuple_name="TTreeName",
    fields=fields,
    variables=variables,
    event_variables=event_variables,
    inputs=mc_data_TES_location)
  \end{lstlisting}
  \end{minipage}
\noindent

\begin{center}
\begin{minipage}{.46\textwidth}
\begin{lstlisting}[
  firstnumber=1,
frame=tlrb, language=Python, caption={\texttt{FunTuple\_Event} instance\vspace{0.05cm}}, label={lst:ftevt}]{Name}
# import FunTuple to run over reconstructed or simulated event
from FunTuple import FunTuple_Event

# define instance of FunTuple
data_tuple = FunTuple_Event(
    name="TDirectoryName",
    tuple_name="TTreeName",
    variables=event_variables)
  \end{lstlisting}
  \end{minipage}
\end{center}
\end{lstfloat}

\subsection{Finding decays in an event}
\label{sec:fields}

Given the distinct event models for reconstructed and simulated events, the \texttt{FunTuple} component employs two separate \texttt{Gaudi} tools for decay identification. Specifically, \texttt{FunTuple\_Particles} relies on the 
\texttt{Gaudi} tool~\cite{GaudiService} \texttt{DecayFinder}~\cite{Barre:2839475}, while \texttt{FunTuple\_MCParticles} utilises the \texttt{MCDecayFinder} tool~\cite{lhcbpackage}. Both of these tools utilise the \texttt{boost} library~\cite{BoostRegex,BoostSpirit} to parse decay descriptors. The names of particles used in the decay descriptor, along with their associated properties, are stored in the \lhcb conditions database (\texttt{CondDB})~\cite{lhcbconditions}, and are retrieved through the \texttt{ParticlePropertySvc}~\cite{GaudiPackage} service.

To isolate a particular decay process within an event and 
select a particle within the decay chain, 
the user is required 
to provide a \texttt{fields} attribute to either the \texttt{FunTuple\_Particles} or the \texttt{FunTuple\_MCParticles} instance.
The \texttt{fields} attribute takes the form of a string dictionary. Here, the \texttt{key} corresponds to the particle alias, serving as a prefix to label the \texttt{TBranch} in the resulting output file. On the other hand, the associated \texttt{value} denotes the decay descriptor employed to filter and select the particles participating in a distinct reconstructed or simulated decay process.
A practical illustration of the \texttt{fields} attribute configuration 
is shown in Listing~\ref{lst:fields}.
    
\begin{lstfloat}
\begin{center}
\begin{minipage}{0.88\textwidth}
\setcounter{lstlisting}{3}
  \begin{lstlisting}[
  firstnumber=1,
  frame=tlrb, language=Python, caption={Example definition of the \texttt{fields} attribute.
    \vspace{0.05cm}
    }, label={lst:fields}]{Name}
# define fields to select decays in an event
# key: alias of the particle used as a prefix to name the TBranch
# value: decay descriptor syntax select particles
fields = {
    "Bplus": "[B+ ->  (J/psi(1S) -> mu+ mu-) [K+]CC ]CC",
    "Jpsi" : "[B+ ->  ^(J/psi(1S) -> mu+ mu-) [K+]CC ]CC",
    "kaons": "[B+ ->  (J/psi(1S) -> mu+ mu-) ^[K+]CC ]CC",
}
  \end{lstlisting}
\end{minipage}
\end{center}
\end{lstfloat}

A correct syntax for the decay descriptor is crucial 
in the selection of the particles within a given decay process.
A straightforward decay descriptor such as \texttt{"B+ -> J/psi(1S) K+"} is 
employed to select all decays of a 
\Bp meson into a \jpsi meson and a \Kp meson.
For the inclusion of charge-conjugate decays, users can encapsulate the decay descriptor in square brackets and append the \texttt{CC} keyword, such as \texttt{"[B+ -> J/psi(1S) K+]CC"}. This syntax covers both \decay{\Bp}{\jpsi\Kp} and 
\decay{\Bm}{\jpsi\Km} decays. Alternatively, the \texttt{[]CC} notation can also be used around an individual particle, e.g., \texttt{"B+ -> J/psi(1S) [K+]CC"}, encompassing both 
\decay{\Bp}{\jpsi\Kp} and 
\decay{\Bp}{\jpsi\Km} decays.\footnote{The charge-violating decays are often reconstructed at \lhcb to serve as proxies for the study of sources of background.}
To target a specific particle within a decay, the caret symbol (\texttt{\^}) is employed. For instance, \texttt{"B+ -> J/psi(1S) \^{}K+"} selects the \Kp meson, while excluding the caret symbol selects the parent particle. In cases of identical particles in the final state, the \texttt{FunTuple} component ensures distinct \CC objects for each identical particle instance. For example, \texttt{"B+ -> \^{}pi+ pi- pi+"} and \texttt{"B+ -> pi+ pi- \^{}pi+"} would choose two distinct instances of a \pip.
In the context of simulations, the \texttt{FunTuple\_MCParticles} component utilises the \texttt{LoKi} decay finder~\cite{LoKiDecayFinder}. 
This finder offers the flexibility to incorporate various arrow types within the decay descriptor syntax~\cite{LoKiDecayFinder,LoKiArrows}. Each arrow type allows users to selectively include simulated particles based on distinct criteria. For instance, the \texttt{=>} arrow type signifies the inclusion of an arbitrary number of additional photons stemming from final state radiation of charged particles when matching the decay.

\subsection{Retrieve event and decay information}
\label{sec:variables}

To extract essential information related to either the event or individual particles within a decay chain, users are required to furnish the \texttt{variables} or \texttt{event\_variables} attribute to \texttt{FunTuple}.
The \texttt{variables} attribute functions as a python dictionary in which the \texttt{key} corresponds to the particle name previously defined in the \texttt{fields} attribute. The corresponding \texttt{value} is an instance of a \texttt{FunctorCollection}, which acts as a collection of \texttt{ThOr} functors, effectively resembling a dictionary itself, with the \texttt{key} representing the variable name and the \texttt{value} denoting a \texttt{ThOr} functor. Within the context of the \texttt{FunTuple} component, these \texttt{ThOr} functors are just-in-time (JIT) compiled and employed on the particle instance to retrieve the desired information.
Notably, a key labelled \texttt{ALL} holds a special significance within the definition of the \texttt{variables}. 
Any \texttt{FunctorCollection} associated with the \texttt{ALL} key is applied to all particles specified in the \texttt{fields} attribute.
In contrast, the \texttt{event\_variables} attribute takes the form of an instance of \texttt{FunctorCollection}. The enclosed \texttt{ThOr} functors are designed to provide information at the event level.
The specifics of how to define the \texttt{variables} and \texttt{event\_variables} attributes are illustrated in Listing~\ref{lst:variables}.

\begin{lstfloat}
\begin{center}
\noindent
\begin{minipage}{.88\textwidth}
  \noindent
  \begin{lstlisting}[
  firstnumber=1,
  frame=tlrb, language=Python, caption={Example definition of the \texttt{variables} and \texttt{event\_variables} attributes.
    \vspace{0.05cm}
    }, label={lst:variables}]{Name}
# import ThOr functor library
import Functors as F
# import the FunctorCollection library
import FunTuple.functorcollections as FC
# import function to get TES location of PVs
from PyConf.reading import get_pvs

# variables for "Bplus" defined in the "fields"
b_vars = FunctorCollection()
# store the flight distance of candidate B relative to the primary vertex that best aligns with the origin of candidate B.
pvs = get_pvs()
b_vars["BPVFD"] = F.BPVFD(pvs)

# variables for "Kaons" defined in the "fields"
kaon_vars = FunctorCollection()
kaon_vars["PT"] = F.PT

# variables for "ALL" particles defined in "fields"
all_vars = FunctorCollection()
all_vars["ETA"] = F.ETA

# define decay-level variables
variables = {
  "Bplus": b_vars,
  "Kaons": kaon_vars,
  "ALL": all_vars,
}

# define event-level variables, 
# for example number of primary vertices
# and add FunctorCollection "SelectionInfo" 
# that stores trigger configuration key (TCK) and 
# decisions of "Hlt1LineName" trigger line
event_variables = FunctorCollection()
event_variables["nPVs"] = F.nPVs
evt_variables += FC.SelectionInfo(selection_type="Hlt2", trigger_lines=["Hlt1LineName"])
  \end{lstlisting}
\end{minipage}
\end{center}
\end{lstfloat}

The \texttt{FunTuple} component utilises the flexibility inherent in \texttt{ThOr} functors to extract a diverse array of information from the event. 
These functors are adaptable enough to accept multiple reconstructed objects as input, enabling the computation of associated information. For instance, consider the functor designed to calculate the flight distance of a particle. To achieve this, the functor takes both the reconstructed primary vertices and the reconstructed particle as input arguments. The usage of this specific functor (\texttt{BPVFD}) is shown in Listing~\ref{lst:variables}.

The functors support all fundamental mathematical operators, including addition, subtraction, multiplication, and division. Additionally, they can undergo transformations such as \texttt{fmath.log(F.CHI2/F.NDOF)}, which, when applied to a reconstructed track, yields the track's $\chi^2$ per degree of freedom.
Furthermore, the output from one \texttt{ThOr} functor can be passed as input to other \texttt{ThOr} functors through a mechanism known as \textit{composition}.
This proves particularly advantageous when users seek to compute an observable that relies on the outcomes of other observables.
All these functionalities are harnessed to provide users with an range of observables via a pre-defined \texttt{FunctorCollection} instance,
which is intended for use in conjunction with \texttt{FunTuple}. 
An illustrative example is the \texttt{SelectionInfo} collection, which gathers the functors employed to store the trigger configuration key (TCK) 
and the event's trigger line decision. Listing~\ref{lst:selectioninfo} outlines the definition of this collection, with its application showcased in Listing~\ref{lst:variables}.

In this listing, the \texttt{SelectionInfo} collection is designed to take two main inputs: 
the type of selection, which can be any of the three stages (\texttt{Hlt1}, \texttt{Hlt2}, or \texttt{Sprucing}), and a list of trigger or Sprucing lines.
In response, it generates a \texttt{FunctorCollection} that incorporates two 
functors: \texttt{F.TCK} for storing TCK information and \texttt{F.DECISION} for storing the 
trigger decision of the specified selection line. 
Such collections do not expose the users to the technical intricacies involved in retrieving the requested information.
In this particular case, the involved functors require the \texttt{DecReport} object, 
which is obtained from the \texttt{DaVinci} framework via the \texttt{get\_decreports} function.
Furthermore, users maintain the flexibility to add, merge or remove observables within these collections, enabling them to create their customised collections. 
Multiple collections have been developed and continue to be actively expanded, accompanied by relevant unit tests within the \texttt{DaVinci} framework.

\begin{lstfloat}
\begin{center}
\noindent
\begin{minipage}{.88\textwidth}
  \noindent
  \begin{lstlisting}[
  firstnumber=1,
  frame=tlrb, language=Python, caption={Definition of the \texttt{SelectionInfo} collection.
    \vspace{0.2cm}
    }, label={lst:selectioninfo}]{Name}
from GaudiConf.LbExec import HltSourceID
import Functors as F 
from PyConf.reading import get_decreports

def SelectionInfo(*, 
  selection_type: HltSourceID, 
  trigger_lines: list[str]) -> FunctorCollection:
  """ 
  Event-level collection for tupling trigger/Sprucing information.

  Args:
      selection_type (HltSourceID): Name of the selection type i.e. "Hlt1" or "Hlt2" or "Spruce". Used as branch name prefix when tupling and as source ID to get decision reports.
      trigger_lines (list(str)): List of line names for which the decision is requested.
  
  Returns:
      FunctorCollection: Collection of functors to tuple trigger/Sprucing information.
  """

  # get selection type
  selection_type = HltSourceID(selection_type)

  # get decreports
  dec_report = get_decreports(selection_type)

  # check that the code ends with decision
  trigger_lines = [s + "Decision" if not s.endswith("Decision") else s for s in trigger_lines]

  # create trigger info dictionary
  trigger_info = FunctorCollection({
      selection_type.name + "_TCK": F.TCK(dec_report),
      l: F.DECISION(dec_report, l) for l in trigger_lines
  })
  return trigger_info
  \end{lstlisting}
\end{minipage}
\end{center}
\end{lstfloat}

\subsection{Writing of retrieved information}
\label{sec:output}

The \texttt{ThOr} functors, utilised for retrieving reconstructed or truth-level information, are capable of encapsulating data in a diverse range of formats. These functors 
can return basic \cpp types, but they can also yield complex objects pertaining 
to the \lhcb software framework.
Subsequently, the extracted information is recorded within the ROOT file, where each \texttt{TBranch} corresponds to a scalar or an array of basic \cpp types.
\texttt{FunTuple} accommodates diverse data object types returned by \texttt{ThOr} functors. An illustrative example is the functor \texttt{F.STATE}, which retrieves the complete state of a reconstructed track i.e. instance of \texttt{LHCb::State}, which includes information on track position, charge, momentum, track slopes, and the associated covariance matrix. \texttt{FunTuple} processes this returned class instance, enabling the writing of multiple observables into the ROOT file from a single functor.
In this context, \texttt{FunTuple} supports various variable types, and the list is rapidly expanding. 
These include three-vectors, four-vectors, \texttt{SIMD} versions of arrays, matrices of both symmetric and non-symmetric nature with arbitrary dimensions, containers spanning arbitrary dimensions, various enumerations e.g. vertex type,
track state, as well as \texttt{std::optional<T>} constructs and \texttt{std::map<std::string, T>} structures, where \texttt{T} represents any of the supported types. Additionally, extending support for other custom classes is remarkably straightforward.

As of the preparation of this document, the \texttt{FunTuple} component utilises the \texttt{GaudiTupleAlg} tool~\cite{GaudiPackage}, which registers an entry in the ROOT file in a thread-safe manner. However, this tool does not provide full support for various complex data objects returned by \texttt{ThOr} functors; such support is exclusively offered by \texttt{FunTuple}. 
The transition to ROOT's \texttt{RNTuple} is planned for the future with subsequent retirement of the \texttt{GaudiTupleAlg} tool.

\subsection{Test suite, examples and performance}
\label{sec:tests_examples}

\texttt{FunTuple} includes an extensive set of examples and tutorials for users, along with a dedicated test suite based on \texttt{pytest}~\cite{pytest}. 
%The test suite has also been developed for the decay finder tool; refer to Ref.~\cite{Barre:2839475} for more details. 
Both unit tests and ``physics tests" are crafted to assess various functionalities of the component, ensuring its reliability. 
Additionally, an application test accompanies each example job run in continuous integration, serving to guarantee correct functionality consistently.

Comprising just over 100 unit tests and some 40 ``physics tests", the test suite currently in place evaluates various aspects of the FunTuple behaviour.
These include checking for appropriate error messages in case of incorrect configurations, ensuring correct output with specified settings, validating expected numbers written to the ROOT file, testing the behaviour of \texttt{FunctorCollections}, assessing the output of FunTuple when run with different event models, and more.
The test coverage for both \texttt{FunTuple} and the decay finder stands at an impressive 100\%.

While a comprehensive performance analysis of \texttt{FunTuple} is not the focus of this paper, a brief overview is provided. In offline analysis, computing hundreds of observables is common. Recording 740 observables using ThOr functors for 1000 events takes 3 minutes, with JIT compilation of about 200 functors taking 84 seconds. Post-compilation, a functor cache is created, reducing overhead in both online and offline data processing. The Python front-end of \texttt{FunTuple} assists in early error detection in configurations, and the performance impact from combining C++/Python is minimal relative to functor execution time.

\section{Interface with other \texttt{Gaudi} algorithms}
\label{sec:otheralgs}

In the \lhcb framework, the execution of multiple algorithms within the offline data processing pipeline is a common necessity. 
Notable examples of such algorithms encompass the \texttt{DecayTreeFitter}\cite{Hulsbergen:2005pu}, which fits complete decay chains with optional primary vertex constraints or mass constraints on intermediary states; 
the \texttt{MCTruthAndBkgCatAlg} algorithm~\cite{Rec}, which is used to extract truth-level information from reconstructed objects in simulations; 
the \texttt{ParticleCombiner} algorithm~\cite{Rec}, for combining basic particles into composite entities; among others. 
These algorithms can be employed in conjunction with the \texttt{FunTuple} component to process and store data.
A practical illustration of \texttt{FunTuple} in synergy with \texttt{DecayTreeFitter} and \texttt{MCTruthAndBkgCat} is presented in Listing~\ref{lst:ftdtf}.

In this listing, the \texttt{DecayTreeFitter} and \texttt{MCTruthAndBkgCat} algorithms operate on reconstructed \decay{\Bp}{\jpsi\Kp} decays. 
Under the hood, both algorithms construct a relation table linking the reconstructed object with a related object that holds pertinent information. 
For \texttt{MCTruthAndBkgCat}, the related object is the associated simulation object, harbouring truth-level information; conversely, for \texttt{DecayTreeFitter}, the related object corresponds to the output of the decay tree fitting process. 
To extract the relevant information, the reconstructed object is mapped to the related object, and the \texttt{ThOr} functor is applied to the related object. 
This entire process is executed within the \texttt{\_\_call\_\_} method of both the \texttt{MCTruthAndBkgCat} and \texttt{DecayTreeFitter} algorithms.
For example, in Listing~\ref{lst:ftdtf}, calling \texttt{MCTRUTH(F.FOURMOMENTUM)}
establishes a mapping between the reconstructed 
\decay{\Bp}{\jpsi\Kp}
decay and the corresponding simulation object. Subsequently, the \texttt{F.FOURMOMENTUM} functor is employed on the simulation object to retrieve the true four-momentum of the \texttt{B+} meson. 
A similar approach is followed for the \texttt{DTF(F.FOURMOMENTUM)}, with the distinction that the four-momentum of the \Bp meson is stored following the decay tree fit, incorporating mass constraint on the \jpsi meson and primary vertex constraint.

The interaction between \texttt{FunTuple} and other \texttt{Gaudi} algorithms is fortified by a fail-safe mechanism. 
When either of the algorithms encounters failure, such as the absence of corresponding truth-level information or unsuccessful decay tree fitting, the \texttt{ThOr} functors and \texttt{FunTuple} are equipped to handle the situation. 
If the \texttt{ThOr} functor returns data of floating-point type, the \texttt{FunTuple} component automatically records Not a Number (\texttt{NaN}) in the ROOT file. Conversely, if the \texttt{ThOr} functor returns an integral type, the invalid value needs to be explicitly defined using the \texttt{F.VALUE\_OR} functor, exemplified in Listing~\ref{lst:ftdtf}. 

\begin{lstfloat}
  \begin{center}
  \noindent
  \begin{minipage}{.88\textwidth}
    \noindent
    \begin{lstlisting}[
  firstnumber=1,
    frame=tlrb, language=Python, caption={
      Usage of truth-matching (\texttt{MCTruthAndBkgCat})
       and decay tree fitting (\texttt{DecayTreeFitter})
       algorithms 
       in conjunction with \texttt{FunTuple}.
       Note that the FunTuple definition 
       shown in the Listing~\ref{lst:ftreco} does not change.
       \vspace{0.2cm}
       },
      label={lst:ftdtf}]{Name}
from DecayTreeFitter import DecayTreeFitter
from DaVinciMCTools import MCTruthAndBkgCat
import Functors as F
from PyConf.reading import get_pvs

# get the TES location of the input data with 
# reconstructed "B+ -> J/psi(1S) K+" decays
input_data = get_particles(f"/Event/HLT2/BToJpsiK/Particles")

#get the reconstructed pvs
pvs = get_pvs()

# define an instance of MCTruthAndBkgCat algorithm for truth-matching.
# Arguments include:
# - name: User-specifed name 
# - input_data: TES location of the input data
MCTRUTH = MCTruthAndBkgCat(name="MCTRUTH", input_data=input_data)

# define an instance of DecayTreeFitter for fitting the decay chain
# Arguments include:
# - name: User-specifed name
# - (optional) mass_constraint: Mass constraint on intermediate state (in this instance J/psi(1S))
# - (optional) input_pvs: TES location of reconstructed primary vertices to apply primary vertex constraint
# - input_data: TES location of the input data
DTF = DecayTreeFitter(name="DTF", mass_constraints=["J/psi(1S)"], input_pvs=pvs, input_data=input_data)

# define the B candidate variables to be passed to FunTuple
# Note: The "F.VALUE_OR" functor specifies an invalid value to be written to ROOT file in the case of no corresponding truth-level information. For functors returning floating point types such as components of F.FOURMOMENTUM, this is automatically chosen to be "NaN" by FunTuple
b_vars = FunctorCollection()
# add truth-level information
b_vars["TRUE_ID"] = F.VALUE_OR(0) @ MCTRUTH(F.PARTICLE_ID)
b_vars["TRUE_FOURMOM"] = MCTRUTH(F.FOURMOMENTUM)
# add decay tree fitter information
b_vars["DTF_FOURMOM"] = DTF(F.FOURMOMENTUM)
  \end{lstlisting}
  \end{minipage}
  \end{center}
\end{lstfloat}

\section{Summary and conclusions}
\label{sec:conclusions}

\begin{figure}[!hpt]
\centering
\includegraphics[width=\textwidth]{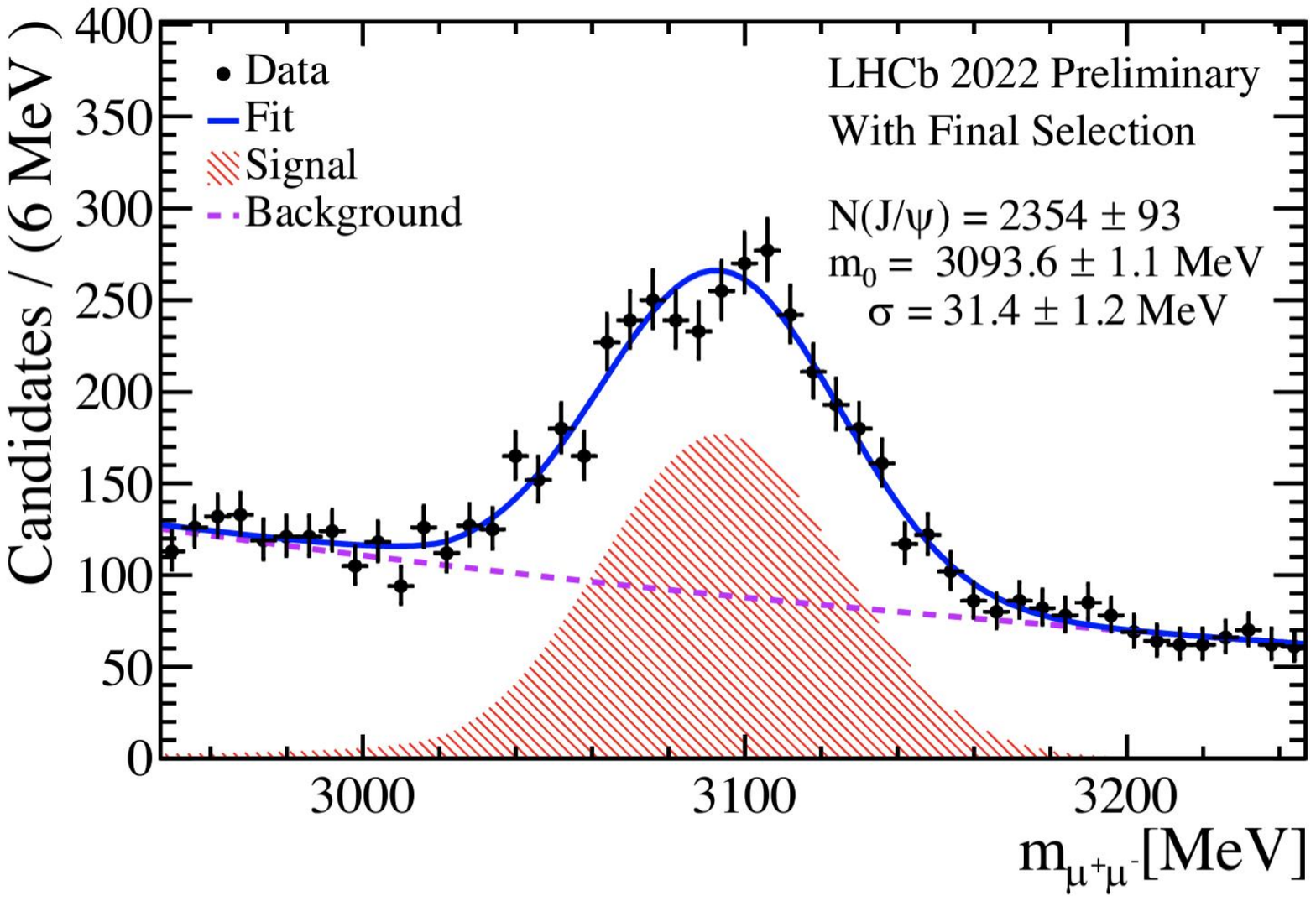}
\caption{Invariant mass of the (\mun\mup) system showing the \jpsi peak
for \lhcb data collected during the current Run 3 commissioning data taking period
in 2022~\cite{LHCBFIGURE2023015}.}
\label{fig:jpsi}
\end{figure}

This paper introduces the \texttt{FunTuple} component, designed to support offline data processing for the \lhcb experiment during the current Run 3 and subsequent runs.
Its primary purpose is to facilitate the storage of experiment-acquired data in the \texttt{ROOT} format, optimising it for subsequent offline analysis. 
Currently, the component plays a vital role in various early measurement analyses of \lhcb data collected during the current Run 3 data taking period. An example of the processed data using \texttt{FunTuple} 
is displayed in Fig.~\ref{fig:jpsi}, showcasing the reconstructed mass of the 
 \decay{\jpsi}{\mun\mup} decay from \lhcb data gathered in 2022 during commissioning~\cite{LHCBFIGURE2023015}. 
The figure shows the signal \decay{\jpsi}{\mun\mup} component in red filled histogram
and the background component in dotted purple line. 
The background component involves random combinations of muons from different part of the event.
The total fit component, composed of both signal and background, is shown in solid blue line and the data 
points are shown in black dots.
The number of signal candidates is estimated to be $N_{\jpsi} = 2354 \pm 93$ with 
mass $m_0 = 3093.6 \pm 0.2$~\mevcc and width $\sigma = 9.1 \pm 0.2$~\mevcc to be consistent with the known \jpsi mass and width~\cite{PDG2020}.

Furthermore, the \texttt{FunTuple} component is built upon the \texttt{Gaudi} functional framework, and it offers a user-friendly \texttt{Python} interface.
Its templated design enables it to accommodate various types of input data, including reconstructed and simulated events, and it supports the processing of both event-level and decay-level information.
Additionally, this templated design allows the component to support new event models, based 
on \textit{SoA} data structure in the future, facilitating vectorised processing of data.
Of particular importance is  its ability to ensure equivalence between trigger-computed observables and those subjected to offline analysis. This achievement is made possible through the integration of the \texttt{ThOr} functors, adept at computing topological and kinematic observables.
Users also have substantial flexibility, enabling them to personalise the range of observables stored within the \texttt{ROOT} file.
The component is also thoroughly validated through a series of unit-tests and \texttt{pytest} tests to ensure its reliability. 
In conclusion, the unique attributes of the \texttt{FunTuple} component establish it as a robust tool for offline data processing at 
the \lhcb experiment making it essential for Run 3 and beyond.

%% Do not include this in any draft (just for information in the template)
%%\input{acknowledgements_intro}
%% Comment this in for paper drafts; do not include this in analysis note, conference and figure reports
\clearpage
\section*{Acknowledgements}
We extend our sincere appreciation to our collaborators in the Data Processing and Analysis (DPA) project for their insightful discussions, input, and unwavering support throughout the work.
We are particularly grateful to Maurizio Martinelli for his work in documenting examples pertaining to the~\texttt{FunTuple}, and to Davide Fazzini for his contribution in developing diverse unit tests for the component.
We acknowledge Sascha Stahl for his tests aimed at optimising the components's speed.
Our appreciation also extends to the members of the Real Time Analysis (RTA) project for their feedback and suggestions on \texttt{ThOr} functor usage.
Additionally, we extend a special thank-you to Christoph Hasse for his contributions to the development of the \texttt{composition} mechanism, which has enhanced the flexibility of using \texttt{ThOr} functors for offline processing.
We also convey our gratitude to the members of the Early Measurement Task Force (EMTF) for Run 3 for their rigorous stress-testing, invaluable feedback, and ongoing work in expanding the \texttt{FunctorCollection} library within the \texttt{DaVinci} framework.
This work received essential support from the Forschungskredit of the University of Zurich under grant number FK-21-129 and the Swiss National Science Foundation under contract 204238.

%\input{supplementary}
%
%\input{appendix}
%
%% This should be taken out in the final paper
%\input{supplementary-app}

\addcontentsline{toc}{section}{References}
%\setboolean{inbibliography}{true}
\bibliographystyle{LHCb}
\bibliography{main}

% \newpage
% \input{Authorship_LHCb-PAPER-20YY-nnn}

% The author list for journal publications is generated from the
% Membership Database shortly after 'approval to go to paper' has been
% given.  It is available at \url{https://lbfence.cern.ch/membership/authorship}
% and will be sent to you by email shortly after a paper number
% has been assigned.  
% The author list should be included in the draft used for 
% first and second circulation, to allow new members of the collaboration to verify
% that they have been included correctly. Occasionally a misspelled
% name is corrected, or associated institutions become full members.
% Therefore an updated author list will be sent to you after the final
% EB review of the paper.  In case line numbering doesn't work well
% after including the authorlist, try moving the \verb!\bigskip! after
% the last author to a separate line.

% The authorship for Conference Reports should be ``The LHCb
% collaboration'', with a footnote giving the name(s) of the contact
% author(s), but without the full list of collaboration names.

% The authorship for Figure Reports should be ``The LHCb
% collaboration'', with no contact author and without the full list 
% of collaboration names.

\end{document}